\documentclass[floatfix,twocolumn,showpacs,amsmath,amssymb]{revtex4-1}
\usepackage{graphicx}
\usepackage{epsfig}
\usepackage{subfigure}
\usepackage{dcolumn}
\usepackage{bm}
\setcitestyle{super}

\begin{document}

\title{ Band gap engineering in finite elongated graphene nanoribbon heterojunctions: Tight-binding model }
\author{Benjamin O. Tayo}
 \affiliation{Physics Department, Pittsburg State University, Pittsburg, KS 66762, USA}

\begin{abstract}
 A simple model based on the divide and conquer rule and tight-binding (TB) approximation is employed for studying the role of finite size effect on the electronic properties of elongated graphene nanoribbon (GNR) heterojunctions. In our model, the GNR heterojunction is divided into three parts: a left (L) part, middle (M) part, and right (R) part. The left part is a GNR of width $W_{L}$, the middle part is a GNR of width $W_{M}$, and the right part is a GNR of width $W_{R}$. We assume that the left and right parts of the GNR heterojunction interact with the middle part only. Under this approximation, the Hamiltonian of the system can be expressed as a block tridiagonal matrix. The matrix elements of the tridiagonal matrix are computed using real space nearest neighbor orthogonal TB approximation. The electronic structure of the GNR heterojunction is analyzed by computing the density of states. We demonstrate that for heterojunctions for which $W_{L} = W_{R}$, the band gap of the system can be tuned continuously by varying the length of the middle part, thus providing a new approach to band gap engineering in GNRs. Our TB results were compared with calculations employing divide and conquer rule in combination with density functional theory (DFT) and were found to agree nicely.
\end{abstract}
\maketitle
\section{Introduction}
Graphene is a two-dimensional (2D) allotrope of carbon with excellent electronic and mechanical properties, making it suitable for multiple applications in nanoscale electronics and nanophotonics.\cite{Novoselovone, Novoselovtwo} A major deficiency in graphene's properties is the absence of a band gap rendering it impossible for use in switching circuits.\cite{Geim} Several approaches have been used to induce a band gap in graphene such as electrically gated bilayer graphene,\cite{Castro,Zhang,McCann} substrate induced band gap,\cite{Zhou, Gionannetti} or isoelectronic codoping with boron and nitrogen.\cite{Lei} Recently, it has become possible to engineer the band gap of graphene by etching or patterning along a given direction to produce ultra narrow quasi one-dimensional (1D) nano sheets referred to as graphene nanoribbons (GNRs)\cite{Hanone, Hantwo, Todd} with remarkable properties such as width-dependent tunable band gaps, exciton-dominated optical spectra, and room temperature ballistic transport.\cite{exciton,Baringhaus,Palacios} As quasi 1D materials, GNRs are extremely sensitive to their surrounding conditions, which provides a route for manipulating their electronic properties. Additionally, other factors such as finite size effect,\cite{Hod,Nakada} edge effect,\cite{Yson,Lee,Sodi,Gorjizadeh,Simbeck,Wang} and the presence of strain\cite{Xihong,Li,Lu, benTayo} could be used to effectively tune the electronic properties GNRs.

As exciting as the properties of GNRs are, a major bottleneck in exploiting their properties is the difficulty of synthesizing high quality GNRs having specific widths and edges. The most common method for synthesizing GNRs is lithographic patterning out of larger 2D graphene sheets. Such a top-down approach lacks precision and often results in GNRs with random shapes, rough edges and poor electronic quality. Recently, a group in Berkeley Lab and UC Berkeley have successfully demonstrated a precise method for synthesizing high quality GNRs with controlled structure via bottom-up molecular engineering\cite{Crommie}. They successfully synthesized ultra-thin graphene nanoribbon heterojunctions with non-uniform width, providing a new approach to band gap engineering. Motivated by such studies and the potential of molecular bottom-up engineering in producing ultrathin GNR heterojunctions with controlled widths and finite lengths, we decided to carry out detailed studies on the electronic properties of finite length ultrathin GNR heterojunctions using TB approximation. TB approximation is a very simple, reliable, and computationally efficient method for electronic structure calculations and has been used successfully for modelling the electronic, optical and transport properties of carbon-based systems.\cite{Yson,benTayo,Saito,orlikowshi,Yongqiang}

In this paper, we present a model based on the divide and conquer rule\cite{divide} and TB approximation for studying the role of finite size effect on the electronic properties of elongated GNR heterojunctions. In our model, the GNR heterojunction is divided into a left (L) part, middle (M) part, and right (R) part. The left part is a GNR of width $W_{L}$, the middle part is a GNR of width $W_{M}$, and the right part is a GNR of width $W_{R}$. We assume that the left and right parts of the GNR heterojunction interacts with the middle part only so that there is no interaction between the left part and the right part. Under this approximation, the total Hamiltonian of the system can be expressed as a block tridiagonal matrix. The individual matrix elements of the total Hamiltonian operator describing the left, middle and right parts, as well as the coupling matrix between right and middle and left and middle can readily be computed using the real space nearest neighbor orthogonal TB approximation. The electronic structure of the GNR heterojunction is analyzed by computing the density of states. We demonstrate that for heterojunctions for which $W_{L} = W_{R}$, the band gap of the system can be tuned continuously by changing the length of the middle part, thus providing useful information for experimental developments. Because our model is based on the TB approximation, significant computational efficiency is gained without compromising accuracy.

This paper is organized as follows: In Sec. \ref{two}, we describe the general formalism. First, we discuss the TB approximation for the infinite periodic system (Sec. \ref{two} A), then we describe how the divide and conquer rule can be applied to describe the properties of finite elongated systems (Sec. \ref{two} B), and in Sec. \ref{two} C we describe the GNR heterojunction using a three-part model and the block tridiagonal matrix. In Sec. \ref{model}, we discuss the model parameters used and describe how the TB real space matrices can be constructed within the divide and conquer rule. In Sec. \ref{results}, we discuss the results. A short summary concludes the paper.
\section{General Formalism}\label{two}
\begin{figure}[t]
\centering
\includegraphics[width=3.2 in]{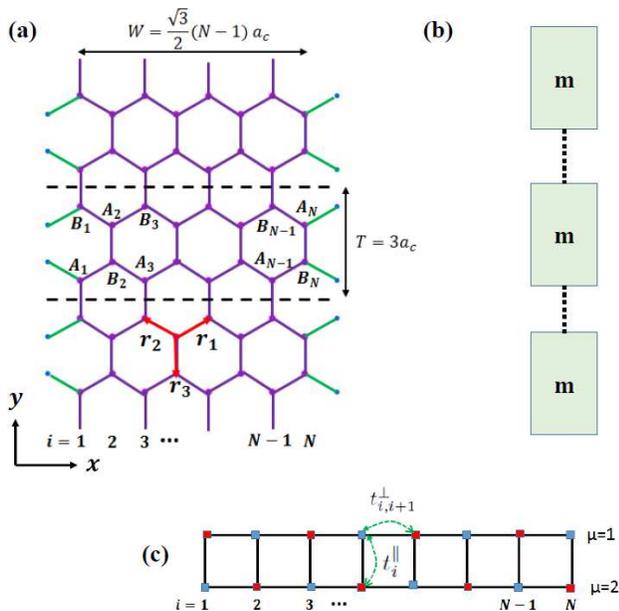}
\caption{(a) Infinite periodic H-passivated AGNR showing the number of dimer lines along the width of the ribbon. (b) Schematic representation of the finite extended GNR, constructed by a replication of the repeating unit ($m$). $m$ represents the unit cell of an N-AGNR. (c) Two-leg ladder with $N$ rings representing the equivalent TB Hamiltonian of the periodic system at the $\Gamma$ point. $t^{\parallel}_{i} $ and $t^{\perp}_{i,i+1}$ are model parameters (hopping integrals) that are used in constructing the TB Hamiltonian matrix.}
\label{AGNR structure}
\end{figure}
We consider an infinite periodic armchair GNR (AGNR) of width $W = {\sqrt{3}\over 2 }(N-1)a_c$ and translation period  $T = 3a_c$, where $N$ is the  number of dimer lines and $a_c \sim 1.43~\mathrm{\AA}$ the unstrained carbon to carbon (C-C) bond length at the center of the GNR (see Fig. \ref{AGNR structure}(a)). Since the width of an AGNR is specified by the number of dimer lines along the ribbon, we will use the notation N-AGNR to refer to an AGNR with N dimer lines along the ribbon. The unit cell of an N-AGNR contains $N$ A-type atoms and $N$ B-type atoms, as shown in Fig. \ref{AGNR structure}. The dangling $\sigma$-bonds at the edges are passivated by H atoms (or other atoms/groups like O and OH). Consequently, the C-C bond length at the edges is shortened to $(1-\delta_{c})a_c$, where $\delta_{c}$ ($3$ to $5 \%$) is the compressive strain on the edge C-C bond due to H passivation\cite{benTayo, Hosoya,Fujita}. Additionally,  N-AGNRs can be classified into three distinct families $N = 3p, 3p+1 , 3p+2$, where $p$ is a positive integer and their electronic properties are known to exhibit distinct family splitting.\cite{Wakabayashi,Ezawa,Brey,Sasaki,Abanin} For the finite extended AGNR (Fig. \ref{AGNR structure}(b)), we introduce an integer length parameter $N_{L}$ such that the length $L$ of the system is $L = N_{L} T$. We will define an N-M GNR heterojunction as a heterojunction formed between an N-AGNR and an M-AGNR.
\subsection{Infinite Periodic GNRs}
The electronic states of infinite periodic (p) GNRs are expressed in terms of the axial momentum ($k$) and the lateral momentum ($k_n$), where $n$ in an integer describing the quantization of the component of electron's momentum along the width of the ribbon. AGNRs are semiconductors with a direct band gap at the $\Gamma$ point. At $k=0$, the TB  Hamiltonian for an AGNR reduces to a two-leg ladder lattice system \cite{Yson}, as shown in Fig. \ref{AGNR structure} (c). The Hamiltonian of this simpler model $\mathcal{H}_p$ (where p stands for periodic) reduces to
\begin{eqnarray}\label{lattice model hamiltonian}
\mathcal{H}_{p} &=& \sum_{i=1}^{N}\sum_{\mu =1}^{2} \varepsilon_{\mu,i}a^{\dagger}_{\mu,i}a_{\mu,i}  -\sum_{i=1}^{N-1} \sum_{\mu =1}^{2} t^{\bot}_{i,i+1}(a^{\dagger}_{\mu,i+1}a_{\mu,i} + \nonumber \\
&& h.c.)-\sum_{i=1}^{N} t^{\|}_{i}(a^{\dagger}_{1,i}a_{2,i} + h.c.),
\end{eqnarray}
where $({i,\mu})$ denote a site, $\varepsilon_{\mu,i}$ site energies, $t^{\bot}_{i,i+1}$ and  $t^{\|}_{i}$ the
nearest neighbor hopping integrals within each leg and between the legs respectively, and $a_{\mu,i}$ the annihilation
operator of $\pi$-electrons on the $i$-th site of the $\mu$-th leg. We remark here that in this model, the electronic properties of GNRs are sensitive only to the three parameters: the site energies $\varepsilon_{\mu,i}$, and the nearest neighbor hopping integrals $t^{\bot}_{i,i+1}$ and $t^{\|}_{i}$. These parameters will differ for perfectly terminating, edge passivated
and strained GNRs. In general for $k \neq 0$, $\mathcal{H}_p$ can be expressed in matrix form for the translational invariant system. If we order the basis as $A_1$, $B_2$, $A_3$, $\dots$, $A_{N-1}$, $B_{N}$, and $B_1$, $A_2$ , $B_3$, $\dots$, $B_{N-1}$, $A_N$, then the nearest neighbor Hamiltonian can be split into four $N\times N$ blocks
\begin{eqnarray}\label{lattice hamiltonian block}
\mathcal{H}_{p}(k) = \left(\begin{array}{cc}
                      \mathcal{H}_1 & \mathcal{H}_{12} \\
                     \mathcal{H}_{12}^{\dagger} & \mathcal{H}_{2}
                   \end{array}\right),
\end{eqnarray}
where
\begin{eqnarray}\label{lattice block one}
\mathcal{H}_1 &=& \left(\begin{array}{cccc}
                      \varepsilon_{1,1} & t^{\bot}_{1,2}&0 & \dots \\
                      t^{\bot *}_{1,2}& \varepsilon_{1,2} & t^{\bot}_{2,3}&\dots  \\
                      0&  t^{\bot *}_{2,3} &\varepsilon_{1,3}& \dots \\
                      \dots & \dots & \dots & \dots
                   \end{array}\right), \nonumber\\
 \mathcal{H}_2 &=& \left(\begin{array}{cccc}
                      \varepsilon_{2,1} & t^{\bot}_{1,2}&0 & \dots \\
                      t^{\bot *}_{1,2}& \varepsilon_{2,2} & t^{\bot}_{2,3}&\dots  \\
                      0&  t^{\bot *}_{2,3} &\varepsilon_{2,3}& \dots \\
                      \dots & \dots & \dots & \dots
                   \end{array}\right),\nonumber \\
\mathcal{H}_{12} &=& \left(\begin{array}{cccc}
                      t^{\|}_{1} d_k & 0&0 & \dots \\
                     0&  t^{\|}_{2}  & 0&\dots  \\
                      0&  0 & t^{\|}_{3} d_k& \dots \\
                      \dots & \dots & \dots & \dots
                   \end{array}\right).\nonumber \\
\end{eqnarray}
Here, $d_k = e^{-i kT}$, with $T$ being the lattice constant. The electronic band structure of the AGNR can then be obtained by solving the eigenvalue equation
\begin{eqnarray}\label{lattice model hamiltonian}
\mathcal{H}_{p}(k) \mathbf{\mathcal{C}}_{\lambda n}(k) =  E_{\lambda n}(k)\mathbf{\mathcal{C}}_{\lambda n}(k),
\end{eqnarray}
where $\lambda = c ~(v) $ corresponds to the conduction (valence) band, and $n$ the band index. The coefficients $\mathbf{\mathcal{C}}_{\lambda n}(k)$ are TB wave function amplitudes.
\subsection{Finite Extended GNRs}
The eigenstates of a finite extended GNR are discrete and can be labelled by the quantum number $i$. Finite size effects in GNRs are studied using the divide and conquer rule. In this approach, the finite-length GNR is constructed by a replication of a repeating unit ($m$), as shown in Fig. \ref{AGNR structure} (b). The TB Hamiltonian $\mathcal{H}_{f}$ (where f stands for finite size) of the system is then given by a block-tridiagonal matrix:

\begin{eqnarray}\label{lattice_block_two}
H_{f} =
 \begin{pmatrix}
  H_{m} & H_{m,m} & 0&\cdots & 0 \\
   H_{m,m}^{\dag} & H_{m} &H_{m,m}& \cdots & 0 \\
   0& H_{m,m}^{\dag} & H_{m}&\cdots & 0\\
  \vdots  & \vdots  & \ddots & \ddots &\vdots \\
  0 & 0 & 0&\cdots & H_{m}
 \end{pmatrix}.
\end{eqnarray}
where $H_{m}$ is the TB Hamiltonian of the subunit $m$, and $H_{m,m}$ the coupling between nearest neighbor subunits. The simplicity of our approach lies in the fact that both $H_{m}$ and $H_{m,m}$ are approximated using the nearest neighbor TB approximation. Our TB description makes calculations easy to compute compared to other methods that use density functional theory (DFT) calculations to compute the matrices $H_{m}$ and $H_{m,m}$\cite{Hod,divide}.
\subsection{GNR Heterojunctions}
\begin{figure}[t]
\centering
\includegraphics[width=3.0 in]{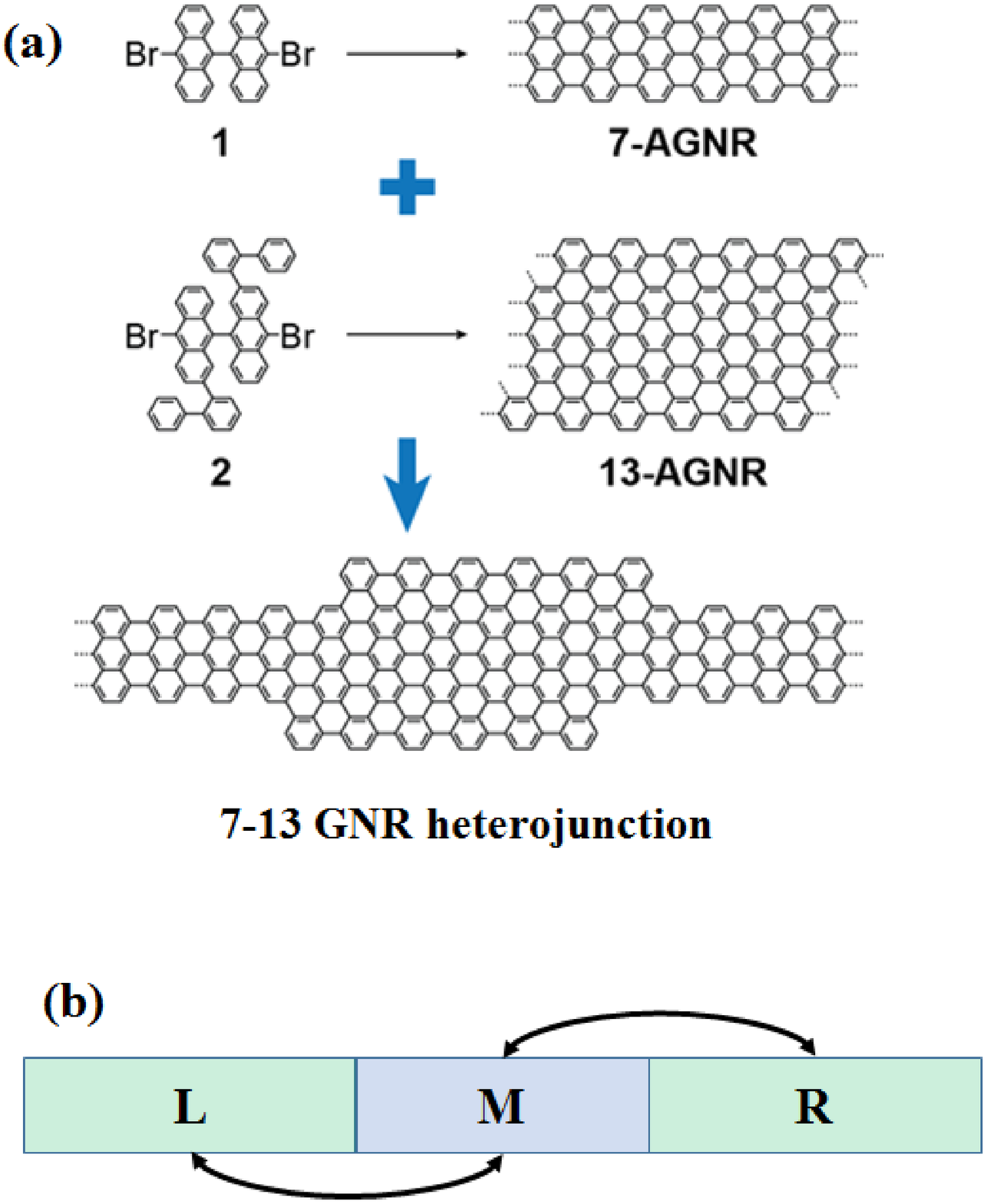}
\caption{(a) Bottom-up synthesis of graphene nanoribbons from molecular building blocks 1 and 2. The resulting ribbon, or heterojunction, has varied widths as a result of different width molecules 1 and 2. Adapted by permission from Macmillan Publishers Ltd: Nature Nanotechnology (Ref. 28), copyright 2015. (b) Schematic representation of the equivalent system consisting of three parts: left (L), middle (M), and right (R). In this case, the left and right parts are infinitely long 7-AGNR, while the middle part is a finite length 13-AGNR.}
\label{hetero_AGNR structure}
\end{figure}
We consider a finite elongated GNR heterojunction formed between GNRs of different widths (Fig. \ref{hetero_AGNR structure}). We assume that the GNR heterojunction can be divided into three parts: the left (L) part which is an N-AGNR, the middle (M) part which is an M-AGNR and the right (R) part which is an N-AGNR. Again using the divide and conquer rule, the TB Hamiltonian $H_{h}$ (where h stands for heterojunction) of the GNR heterojunction can be described by the tridiagonal matrix:
\begin{eqnarray}\label{lattice block three}
\mathcal{H}_{h} &=& \left(\begin{array}{ccc}
                      \mathcal{H}_{f}(L) & \mathcal{H}_{L,M}&0\\
                      \mathcal{H}_{M,L}& \mathcal{H}_{f}(M)& \mathcal{H}_{M,R}\\
                      0 & \mathcal{H}_{R,M}& \mathcal{H}_{f}(R)\\
\end{array}\right).
\end{eqnarray}
where $\mathcal{H}_{f}(L) $ is the finite length TB Hamiltonian matrix of an N-AGNR, $\mathcal{H}_{f}(M) $ is the finite length TB Hamiltonian matrix of an M-AGNR, and $\mathcal{H}_{L,M}$ the coupling matrix between the N-AGNR and M-AGNR forming the heterojunction.

\section{Model Parameters and Computational Details}\label{model}
In this section, we provide some details about the different parameters used in our model and describe how to calculate the block matrices using the TB approximation. For perfectly terminating AGNR, we will set the nearest neighbor C-C TB hopping integral to $ t = 2.7$ eV, a value that has been used to successfully describe the electronic properties of graphene \cite{Reich}. For H-passivated AGNRs, the bond lengths parallel to dimer lines at edges are compressed by about $3.5 \%$ relative to those in the middle of the ribbon. A decrease in C-C bond length will increase overlap of $\pi$ orbitals which leads to an increase in the hopping integral. Likewise, an increase in bond length will result to a decrease in $\pi$ orbital overlap, which accordingly decreases the hopping integral. The analytic expressions for TB matrix elements between carbon atoms as a function of the C-C bond length can be expressed in terms of the Chebyshev polynomials $T_m(x)$  yielding\cite{Porezag}
\begin{eqnarray}\label{carbon to carbon hopping integral}
H_{\pi}^{CC}(r) = \sum_{m=1}^{10} c_{m} T_{m-1}(y) -{c_1 \over 2}, ~~ y = {r - {b+a \over 2} \over {b-a \over 2}}
\end{eqnarray}
where $r \in (a, b)$ is the interatomic distance for C-C interactions, and  $(a, b)$ the range of values over which the expansion is valid. The coefficients $c_m$ and boundaries $a$ and $b$ are tabulated in Ref. 41. Using Eq. (\ref{carbon to carbon hopping integral}), we can show that a $3.5 \%$ compressive strain on the bond length at the edges induces a $12 \%$ increase ($\delta_t$) in the hopping integral . The effect of H-passivation can then be accounted for by setting $t^{\|}_{i} = (1+\delta_t)t $, for $i=1$ and $i=N$, $t^{\|}_{i} = t $  for $i=2, \dots, N-1$, and $t^{\bot}_{i,i+1} = t$  for $\mu =1,2$, $i = 1, \dots N-1$.  Previous studies carried out for edge passivated GNRs have shown that to first-order, the change in onsite energy due to edge passivation does not alter the band gap.\cite{Yson,Wang} We will therefore assume that changes in onsite energies due to H-passivation are negligible. Hence we shall set all the onsite energies at $ \varepsilon_{\mu,i} = 0$ for $\mu = 1,2$ and $n = 1,2, \dots, N$.

Using the approximations given above, we can show that for the finite system, the orthogonal TB nearest neighbor (NN) real space Hamiltonian $H_{m}$ for each subunit is given by
\begin{eqnarray}\label{TB_Hm_matrix}
H^{i,j}_{m} = \begin{cases} t, & \mbox{if}~ i,j~ \mbox{are NN interior bonds}\\ (1+\delta_t)t,& \mbox{if}~ i,j ~\mbox{are NN edge bonds}\\ 0, & \mbox{otherwise}\end{cases}
\end{eqnarray}
where $\delta_{t} \simeq 12 \% $ is the percent increase of the hoping integral due to $3.5 \%$ decrease in C-C bond length at the edges. Using the same arguments and assuming that the coupling between replicating units only involve interior bonds, the real space TB matrix $H_{m,m}$ describing the coupling between adjacent $m$ units is given by
\begin{eqnarray}\label{TB_Hm_matrix}
H^{i,j}_{m,m} = \begin{cases} t, & \mbox{if}~ i,j~ \mbox{are NN interior bonds}\\ 0, & \mbox{otherwise}\end{cases}
\end{eqnarray}

For all the GNRs considered, we will assume the C-C bond length at the center is $a_c = 1.43~\mathrm{\AA}$ and the C-C bond length at the edges is $(1-\delta_c)a_c$, with $\delta_c = 3.5 \%$. For electronic band structure calculations, we will neglect the change in band structure of the AGNR due to quasiparticle effects. Quasiparticle effects are known to provide significant improvements in the band gap \cite{LiYang, Prezzi}. Our TB treatment however provides excellent qualitative results that capture the properties of GNRs. Quantitatively speaking, the TB approximation yields results that are comparable to results obtained using DFT in the local density approximation (LDA)\cite{LiYang}. The advantage of using TB model lies in the computational simplicity, compared to DFT methods which are very expensive for describing finite elongated extended systems. All computations were carried out using Mathematica software\cite{mathematica}.
\begin{figure}[t]
\centering
\includegraphics[width=3.1 in]{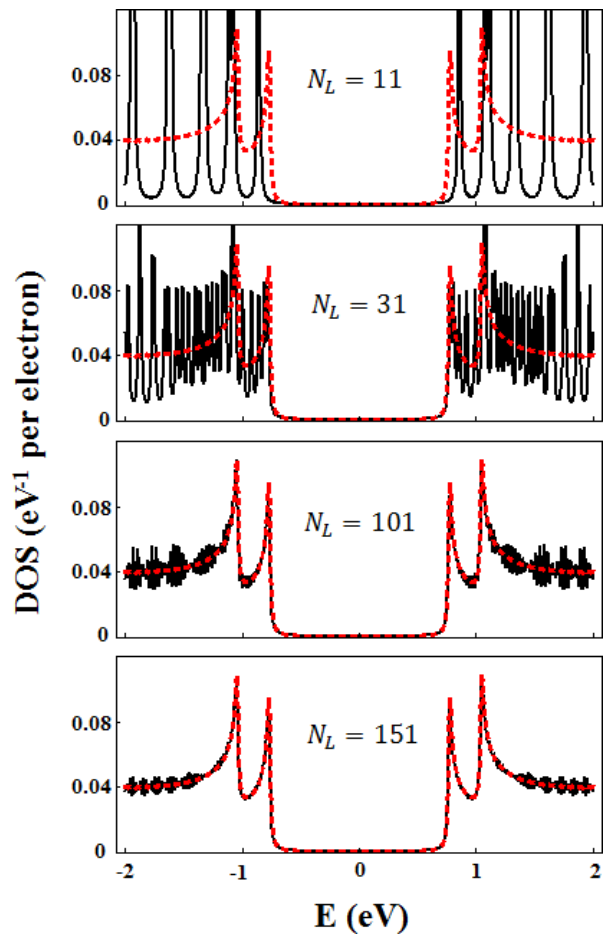}
\caption{DOS of a H-passivated 7-AGNR plotted for different values of the ribbon length $L = N_{L} T$, where $T = 0.43$ nm is the length of the repeating unit used in constructing the finite length system. The DOS is computed with an artificial broadening of 10 meV and compared to the DOS of the infinite periodic system (red dashed curve superimposed on the black curve in each panel). The Fermi energy is set to $E_F =0$. }
\label{7densityofstates}
\end{figure}
\section{Results and Discussion}\label{results}
\begin{figure}[t]
\centering
\includegraphics[width=3.1 in]{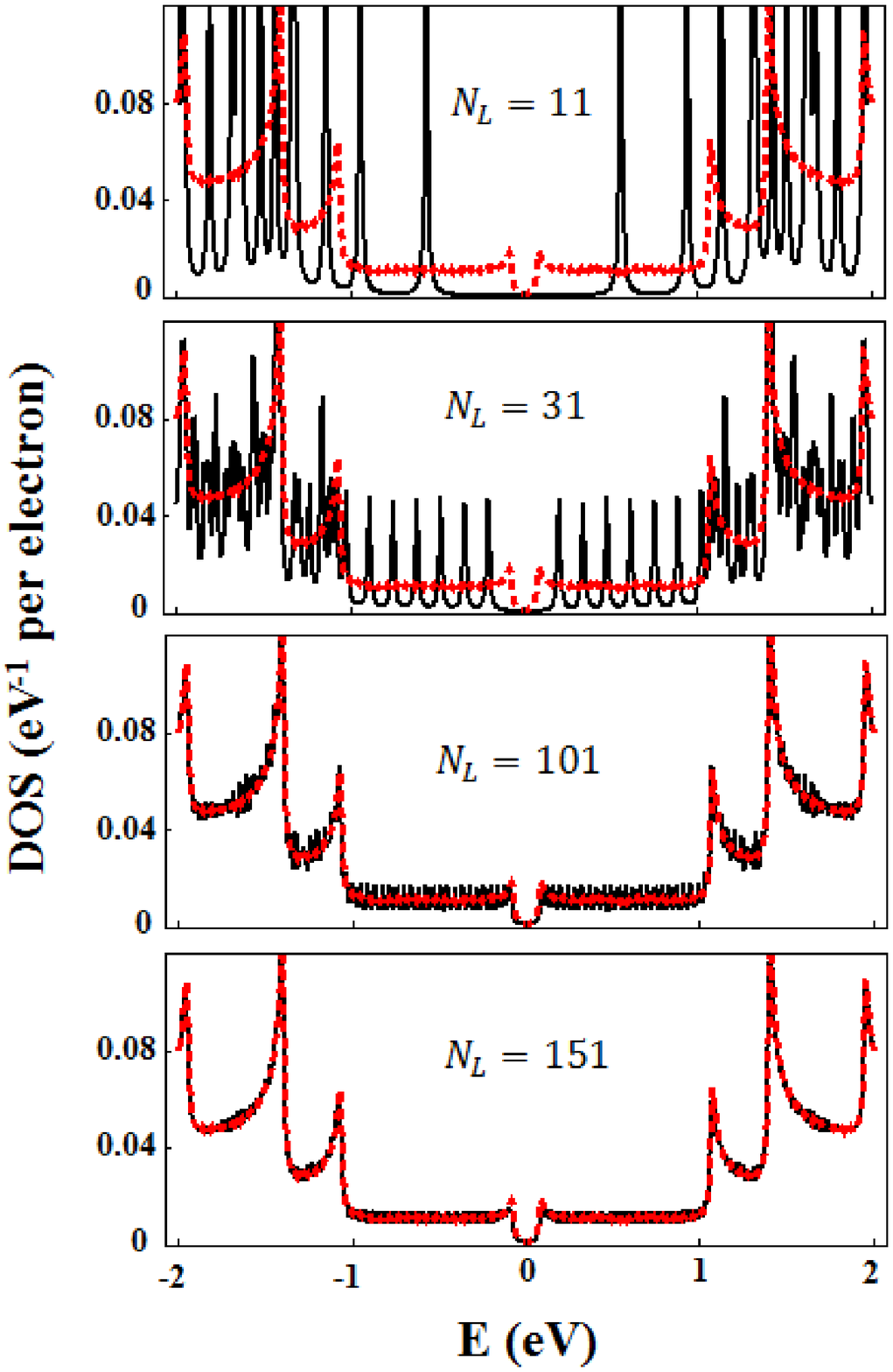}
\caption{DOS of a H-passivated 11-AGNR plotted for different values of the ribbon length $L = N_{L} T$, where $T = 0.43$ nm is the length of the repeating unit used in constructing the finite length system. The DOS is computed with an artificial broadening of 10 meV and compared to the DOS of the infinite periodic system (red dashed curve superimposed on the black curve in each panel). The Fermi energy is set to $E_F =0$. }
\label{11densityofstates}
\end{figure}
\begin{figure}[t]
\centering
\includegraphics[width=3.1 in]{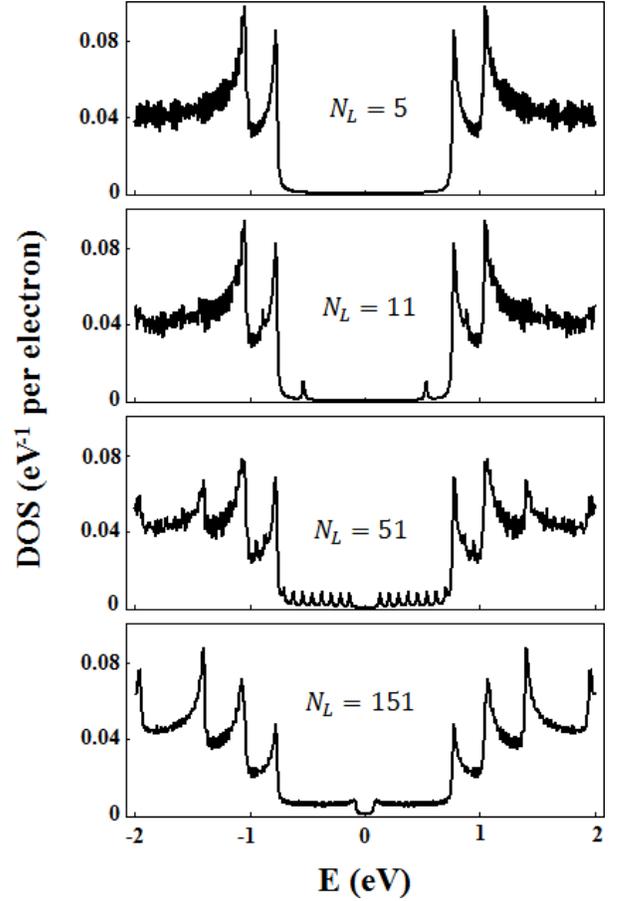}
\caption{DOS of a H-passivated 7-11 GNR heterojunction plotted for different values of the length $L = N_{L} T$ of the 11-AGNR connecting the left and right segments. The DOS is computed with an artificial broadening of 10 meV. The Fermi energy is set to $E_F =0$. As the length of the 11-AGNR is varied, the band gap can be tuned continuously between the band gaps of the 11-AGNR and the 7-AGNR, i.e., between 0.16 eV and 1.54 eV.}
\label{711densityofstates}
\end{figure}
\begin{figure}[t]
\centering
\includegraphics[width=3.1 in]{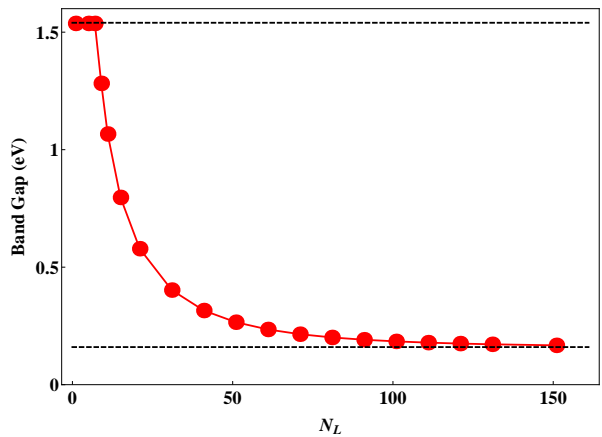}
\caption{Band gap of a H-passivated 7-11 GNR heterojunction plotted for different values of the length $L = N_{L} T$ of the 11-AGNR connecting the left and right segments. Dashed black lines represent the band gaps of the 7-AGNR (1.54 eV) and that of the 11-AGNR (0.16 eV). As the length of the 11-AGNR is increased, the band gap decreases monotonically from 1.54 eV to 0.16 eV. }
\label{gap711}
\end{figure}
We consider a 7-AGNR and an 11-AGNR as choice of materials to be studied. The 7-AGNR belongs to the $3p+1$ family and has a large band gap, while the 11-AGNR belongs to the $3p+2$ family and has a smaller band gap. Our choice of material selection is thus reasonable since our goal is to study band gap engineering in GNR heterojunctions. The 7-11 GNR heterojunction will thus provide a significant range of band gaps lying between the smaller band gap of the 11-AGNR and the larger band gap of the 7-AGNR. Since our goal is to study the properties of finite elongated GNR heterojunctions, we will start by using our TB formalism to study the role of finite size on the electronic properties of the 7-AGNR and the 11-AGNR. Then we will study the effect of finite size on the 7-11 GNR heterojunction and relate its properties to those of the 7-AGNR and an 11-AGNR from which it is made.

We will study the influence of finite size effect on electronic band structure by plotting the density of states (DOS). The finite
temperature DOS per electron of the infinite periodic (p) system $\rho_{p}(E)$ is given by \cite{charlier,Tayo} :
\begin{eqnarray}\label{dosimpurity}
\rho_{p}(E) ={1 \over N_e \Omega} \sum_{n=1}^{N} \sum_{\lambda=v,c}\int_{-\pi/T}^{\pi/T} dk~ \delta[E - E_{\lambda n}(k)],
\end{eqnarray}
where $E_{\lambda n}(k)$ are eigenenergies corresponding to Bloch functions, $N_e$ is the total number of $\pi$ electrons in the GNR, $\Omega = 2\pi/N_LT$ is the length of the 1D reciprocal space for each allowed state, $N_L$ is the length parameter describing the number of repeating units in the N-AGNR of finite length, $L = N_L T$, $T$ being the unit cell length. The density of states of the finite (f) extended system $\rho_{f}(E)$ is given by
\begin{eqnarray}\label{dosfinitesystem}
\rho_{f}(E) ={1 \over N_e } \sum_{i=1}^{N_e} \delta(E - \varepsilon_{i} ),
\end{eqnarray}
where $\varepsilon_{i}$ are eigenenergies of the discrete system. For comparison purposes, both $\rho_{p}(E)$ and $\rho_{f}(E)$ are normalized such that
$$ \int_{-\infty}^{\infty} \rho_{p}(E) dE = \int_{-\infty}^{\infty} \rho_{f}(E) dE =1 .$$
For computational purposes, we replace the Dirac delta function with a Lorentzian with line width $\Gamma = 0.01$ eV. We present the DOS for an energy range of $\pm 2$ eV around the Fermi energy $E_F =0$ for several ribbon lengths $L = N_{L} T$, where $T = 0.43$ nm is the unit cell length.

Figure \ref{7densityofstates} and   \ref{11densityofstates} show the DOS as a function of ribbon length for a hydrogen passivated 7-AGNR and 11-AGNR, respectively. It can be seen that as the length of the system increases, the DOS evolves from a set of irregularly spaced energy levels (typical of finite length molecular systems) to van Hove singularities (VHSs) which is an important characteristic of 1D infinitely periodic systems. When the length of the ribbon exceeds $101 ~T$, finite size effects become negligible and the DOS of the finite system converges to the DOS of the infinite periodic system. Similar results have been obtained using divide and conquer method in combination with DFT\cite{Hod}. The advantage of our method lies in the simplicity of the model and computational efficiency. The band gap of the infinite system can be obtained from the DOS. For the 7-AGNR, the band gap is 1.54 eV while that for the 11-GNR is 0.16 eV. These values agree very well with DFT-LDA band gaps\cite{LiYang}.

Next we consider the 7-11 GNR heterojunction. The left part of the heterojunction is a 7-AGNR of length $L = 101 ~T$, the middle part is an 11-AGNR of variable length $L = N_{L} T$, and the right part is a 7-AGNR of length $L = 101 ~T$. This model allows us to study the properties of the 7-11 GNR heterojunction as a function of the width of the 11-AGNR segment, connecting the left and right parts of the GNR heterojunction. Figure \ref{711densityofstates} shows the DOS of the hydrogen passivated 7-11 GNR heterojunction plotted as a function of the length of the 11-AGNR which forms the middle of the heterojunction. Three important regimes can clearly be seen from the DOS plot. When $N_{L} \leq 7$, the DOS is dominated by the DOS of the 7-AGNR and the band gap of the system is approximately equal to 1.54 eV, which is the band gap of the 7-AGNR. When $7 < N_{L} < 101 $, the DOS contains features from both the 7-AGNR and the 11-AGNR with the band gap varying between the band gap of the 11-AGNR and the 7-AGNR, i.e., between 0.16 and 1.54 eV. When $N_{L} \ge 101$, the DOS of the 7-11 GNR heterojunction is dominated by the DOS of the 11-AGNR and the band gap converges to 0.16 eV. This is a remarkable result and it clearly shows that band gap engineering in GNR can be achieved by synthesizing GNR heterojunctions with varied widths.

In Fig. \ref{gap711}, we plot the band gap of the 7-11 GNR heterojunction for different values of the length $L = N_{L} T$ of the 11-AGNR connecting the left and right segments. The figure shows that by increasing the length of the 11-AGNR, we can tune the bang gap of the material continuously between the band gap of the 7-AGNR (1.54 eV) and the band gap of the 11-AGNR (0.16 eV). When $N_{L} < 7$, the band gap of the 7-11 GNR heterojunction is approximately equal to that of the 7-AGNR (1.54 eV). When $7 \leq N_{L} < 101$, the band gap decreases monotonically from 1.54 eV to 0.16 eV. In this region, the energy gap ($E_{g}$) varies with $N_{L}$ as $E_g =  A + B N_{L}^{-1}$, where $A = 0.06$ eV and $B = 10.6$ eV. The fit is good to within a mean absolute error of 0.008 eV. When $N_{L} \ge 101$, the band gap of the system converges to the band gap of an 11-AGNR (0.16 eV).
\section{Conclusion}
In summary, we have shown that the relevance of finite size effects on the electronic properties of GNRs can well be accounted for by using the divide and conquer rule combined with the TB approximation.  Because our model is based on the TB approximation, significant computational efficiency is gained without compromising accuracy. By computing and analyzing the DOS, we found that for H-passivated AGNRs, the transition from a discrete molecular system to a continuum periodic system occurs at a length of about 43 nm, which is quite consistent with previous results obtained using divide and conquer method in combination with DFT\cite{Hod}. For the 7-11 GNR heterojunction, the DOS and energy band gap plots as a function of the length parameter $N_{L}$ of the middle part reveal three important regimes: when $N_{L} < 7$, the band gap of the 7-11 GNR heterojunction is approximately equal to that of the 7-AGNR (1.54 eV), when $7 \leq N_{L} < 101$, the band gap decreases monotonically from 1.54 eV to 0.16 eV, and when $N_{L} \ge 101$, the band gap of the system converges to the band gap of an 11-AGNR (0.16 eV). Our studies clearly show that band gap engineering in GNRs can be achieved by synthesizing GNR heterojunctions with varied widths, thus providing useful information for experimental developments.

\section*{Acknowledgement}
The author acknowledges support from the department of physics, Pittsburg State University.

\end{document}